\documentclass[sigconf]{acmart}

\AtBeginDocument{%
  \providecommand\BibTeX{{%
    \normalfont B\kern-0.5em{\scshape i\kern-0.25em b}\kern-0.8em\TeX}}}

\copyrightyear{2026}
\acmYear{2026}
\setcopyright{cc}
\setcctype{by}
\acmConference[KDD '26]{Proceedings of the 32nd ACM SIGKDD Conference on Knowledge Discovery and Data Mining V.2}{August 09--13, 2026}{Jeju Island, Republic of Korea}
\acmBooktitle{Proceedings of the 32nd ACM SIGKDD Conference on Knowledge Discovery and Data Mining V.2 (KDD '26), August 09--13, 2026, Jeju Island, Republic of Korea}
\acmDOI{10.1145/3770855.3818497}
\acmISBN{979-8-4007-2259-2/2026/08}

\usepackage{amsmath,amsfonts,bm}

\def\eqref#1{equation~\ref{#1}}

\def\1{\bm{1}}

\DeclareMathAlphabet{\mathsfit}{\encodingdefault}{\sfdefault}{m}{sl}
\SetMathAlphabet{\mathsfit}{bold}{\encodingdefault}{\sfdefault}{bx}{n}

\usepackage[utf8]{inputenc} 
\usepackage[T1]{fontenc}    
\usepackage{hyperref}       
\usepackage{url}            
\usepackage{booktabs}       
\usepackage{amsfonts}       
\usepackage{nicefrac}       
\usepackage{microtype}      
\usepackage{graphicx}
\usepackage{amsmath,amsfonts,amsthm}
\usepackage{array}
\usepackage{subcaption}
\usepackage{bbm}
\usepackage{bm}
\usepackage{multicol}
\usepackage{lipsum}
\usepackage{wrapfig}
\usepackage[ruled]{algorithm2e}
\usepackage{arydshln}
\usepackage{multirow}
\usepackage{enumitem}
\usepackage{siunitx}
\usepackage{dsfont}
\usepackage{xcolor}
\usepackage{tcolorbox}

\newtcolorbox{remarkbox}{
  colback=gray!5,
  colframe=gray!60,
  boxrule=0.5pt,
  arc=2pt,
  left=6pt,
  right=6pt,
  top=6pt,
  bottom=6pt
}

\begin{document}

\title{TransX: Scaling Transformer-based Recommendation via Behavioral and Serving Stream Crossings}


\author{
Da Xu \and
Liyan Fang \and
Divya Venugopalan
}
\affiliation{%
  \institution{LinkedIn}
  \city{Sunnyvale}
  \state{California}
  \country{USA}
}
\email{daxu5180@gmail.com}
\email{lfang@linkedin.com}
\email{dvenugopalan@linkedin.com}

\author{
Sunny Hsu \and
Xukai Wang \and
Rishav Roy Chowdhury 
}
\affiliation{%
  \institution{LinkedIn}
  \city{Sunnyvale}
  \state{California}
  \country{USA}
}
\email{suhsu@linkedin.com}
\email{xuwang@linkedin.com}
\email{richowdhury@linkedin.com}

\author{
Cindy Liang \and
Nishant Satya Lakshmikanth 
}
\affiliation{%
  \institution{LinkedIn}
  \city{Sunnyvale}
  \state{California}
  \country{USA}
}
\email{cliang@linkedin.com}
\email{nlakshmikanth@linkedin.com}

\renewcommand{\shortauthors}{Da Xu et al.}




\begin{abstract}
Modern industrial recommender systems (RecSys) increasingly adopt Transformer-based sequence models, with an emerging paradigm that frames recommendation as next-token prediction over a unified monolithic user sequence. However, collapsing heterogeneous data sources -- such as long-term user behaviors and real-time serving events -- into a single monolithic token stream that obscures their distinct causal roles and temporal characteristics, leading to inefficient modeling and elevated training and serving costs.

We propose TransX, a production-oriented encoder–decoder architecture that reformulates recommendation as a sequence-to-sequence action transduction problem. TransX explicitly decouples behavior-stream modeling from serving-event modeling and conditions next-action decoding on scalable cross-attention between nearline behavior encodings and real-time serving representations. To enable low-latency, high-QPS deployment, TransX is co-designed with an amortized serving strategy that combines incremental behavior encoding with per-request key–value caching, rendering serving latency insensitive to behavior sequence length.

Extensive offline experiments and large-scale online A/B tests on LinkedIn’s recommender systems show that TransX consistently outperforms state-of-the-art DLRMs and sequential baselines, and delivers substantial CTR lift ($+6.0\%$) and conversion gain ($+4.4\%$) while maintaining serving costs comparable to existing production models where our co-designed serving strategy reduces online computation by approximately $80\%$.
\end{abstract}



\maketitle

\section{Introduction}
\label{sec:intro}
Industrial recommender systems (RecSys) are the primary engine of user engagement for online platforms, requiring models that can process billions of events daily with sub-second latency \cite{ricci2021recommender}. For the past decade, Deep Learning Recommendation Models (DLRMs) have dominated this space by leveraging human-engineered features and sparse interactions \cite{zhang2019deep,naumov2019deep}. While effective, DLRMs are fundamentally constrained by their limited ability to ingest ultra-long, heterogeneous behavioral histories, often relying on static aggregations that lose the rich temporal semantics of user intent.

Motivated by the success of Large Language Models (LLMs), recent work has begun to reformulate recommendation via sequence transduction or generative modeling \cite{geng2022recommendation,li2024large,wang2025generative}. In this emerging paradigm, user-system interactions are sequentialized into token streams, and the recommendation task is treated as a sequence transduction problem -- mapping an input sequence to an output sequence of user actions \cite{han2025mtgr,zhang2025onetrans} -- or next-item generation problem \cite{deng2025onerec,zhai2024actions,rajput2023recommender}. This shift allows models to scale their capacity with increased sequence length, tokenization complexity, and compute, reducing the reliance on manual feature engineering in favor of automated, Transformer-based feature interactions.

Despite these advances, many architectures in this emerging paradigm adopt a "monolithic" design that closely mirrors standard LLM training \cite{li2024large,wang2025generative}. Specifically, they collapse heterogeneous data streams — including user historical behaviors and system serving events (e.g., candidate exposures) — into a single unified token sequence. 
While effective in some settings, this design introduces two key limitations in many industrial recommender systems:
\begin{enumerate}[leftmargin=*]
    \item \textbf{Semantic conflation}: it forces the model to treat heterogeneous data generation processes as semantically homogeneous, obscuring the distinct causal and modeling roles of user historical behaviors versus system exposures in recommendation.
    \item \textbf{Computational inefficiency}: collapsing the streams increases sequence length and tokenization complexity, and requires re-encoding the entire behavior prefix for each candidate request, leading to substantial redundant computation.
\end{enumerate}
Our work aims to address these limitations in settings where heterogeneous signals and serving efficiency are both critical.


\paragraph{First part of our solution: architecture innovation} 
To address these challenges, we first introduce TransX, an architecture built on the principle of \textbf{Action Transduction via Stream Crossing}. We observe that recommendation is naturally a dual-stream process: (i) a serving-event stream, in which the system selects and presents a slate of candidate items to the user, and (ii) a behavior stream, which captures the user’s broader activity across the platform, including both responses to recommendations (e.g. \texttt{clicks} or \texttt{likes}) and organic actions (e.g. \texttt{searches} or \texttt{profile views}). 
TransX models a user’s response as the interaction between their long-term behavioral trajectory and the system’s immediate serving slate. And motivated by the distinct generative processes and statistical roles of these streams, it explicitly separates them using an encoder–decoder architecture
The encoder processes the behavior stream to learn latent user intent, while the decoder performs parallel action transduction, mapping the crossing of the behavior stream and the serving slate into a discrete vocabulary of action tokens (e.g. \texttt{click}, \texttt{view}, \texttt{dismiss}).

\paragraph{Second part of our solution: model-infrastructure co-design}
Recognizing that large-scale recommendation in industry is as much about infrastructure as it is about modeling. We co-design TransX with a streaming sequence-to-sequence training pipeline that substantially increases training throughput. By performing the behavior encoding nearline and caching the resulting representations, we shift the bulk of the model’s computational cost away from the critical serving path. At request time, only a lightweight cross-attention "crossing" and action decoding are performed. We also incorporate KV caching and parallel decoding into the design to further reduce serving latency. Together, they allow TransX to scale to 180-day behavior windows while maintaining an online serving cost comparable to traditional DLRMs in LinkedIn's large-scale deployment.

\paragraph{Real-world impact}
To validate TransX's production impact, we conducted extensive experiments on LinkedIn’s social recommendation dataset where TransX consistently outperforms both traditional DLRMs and recent sequence modeling baselines under a similar compute budget. In large-scale online A/B experiments, TransX delivers the largest CTR lift observed in recent years for a major social recommendation application in LinkedIn. 

The key contributions of this paper are summarized as follows:
\begin{itemize}[leftmargin=*]
    \item \textbf{Transduction Framework}: we describe a sequence-to-sequence action transduction formulation that conditions on the behavioral stream and serving stream crossing, and map them to an action token stream.
    \item \textbf{Architectural Innovation}: we propose a grouped multi-query sparse cross-attention mechanism that enables efficient, candidate-specific interaction with long-term user history.
    \item \textbf{Production Scalability}: we detail the model-infrastructure co-design, including nearline amortized encoding and KV-caching, that enables the large-scale deployment of TransX.
    \item \textbf{Empirical Validation}: we provide evidence from LinkedIn’s industrial datasets and online A/B tests, showing record-level CTR gains and favorable serving economics.
\end{itemize}


\section{Related Works}
\label{sec:prelim}
We briefly review representative recommendation models that have demonstrated production impact, focusing on the evolution from DLRM to sequential and generative recommendation.

\paragraph{Conventional DLRM} 
Deep learning recommendation models (DLRM) have become the backbone of many industrial recommender systems. These models learn implicit feature crossings over sparse and dense inputs using deep neural networks.
Deep \& Wide \cite{cheng2016wide} combines linear models with DNNs to balance memorization and generalization, while subsequent architectures such as DeepFM \cite{guo2017deepfm}, DLRM \cite{naumov2019deep}, DCN \cite{wang2017deep}, DCN-V2 \cite{wang2021dcn}, and xDeepFM \cite{lian2018xdeepfm} further improve expressivity through factorization machines, explicit cross networks, or higher-order interaction modules. Despite their effectiveness, conventional DLRM primarily rely on static feature representations and shallow feature-crossing mechanisms, which makes it challenging to capture long-range dependencies and non-stationary user behavior patterns over time.

\paragraph{Sequential recommendation with Transformer}.
With the emergence of sequential modeling, self-attention and Transformer architectures \cite{vaswani2017attention} have been widely adopted to encode user interaction histories. SASRec \cite{kang2018self} models behavior sequences with causal self-attention, while BERT4Rec \cite{sun2019bert4rec} leverages masked token prediction for bidirectional context modeling during training.
Industrial ranking models such as DIN \cite{zhou2018deep} and BST \cite{chen2019behavior} apply attention over user histories and fuse sequence representations with dense and sparse features for downstream prediction.
More recent work, including TransAct and its extensions \cite{xia2023transact,xia2025transact}, unifies sequential attention with multi-behavior user action modeling via Transformer blocks. While these models bridge the gap of DLRM in modeling sequential user behavior, their training and serving  typically operate at relatively modest computational scale, limiting their ability to fully leverage the rapidly growing compute power.

\paragraph{Scaling up sequential and generative recommendation} 
Recent efforts to scale recommendation models largely follow two complementary directions. One line of work increases modeling capacity by expanding feature interaction depth and effective sequence length through customized Transformer architectures, such as SIM \cite{pi2020search}, HiFormer \cite{gui2023hiformer}, InterFormer \cite{zeng2025interformer}, RankMixer \cite{zhu2025rankmixer}, and LONGER \cite{chai2025longer}.
Another line adopts generative recommendation (GR) formulations, reframing recommendation as a monolithic sequence generation or transduction problem to better exploit large-scale training akin to LLM, including Meta's GRM \cite{zhai2024actions}, MTGR \cite{han2025mtgr}, OneTrans \cite{zhang2025onetrans}, and OneRec \cite{deng2025onerec}. TransX differs from these approaches by explicitly exploiting the structural asymmetry between the user behavior stream and the serving event stream. By decoupling these processes, TransX enables Transformer-based recommendation to scale significantly while maintaining cost efficiency under strict latency SLAs.

\section{Preliminaries}
\label{sec:analysis}

\begin{table}[htb]
    \centering
    \resizebox{\columnwidth}{!}{
    \begin{tabular}{|p{3.5cm}|p{6.3cm}|} 
    \hline
        $u$ & User index \\ \hline
        $j$ & Candidate index within a serving event \\ \hline
        $L$, $d$ & Length of user behavior stream and model hidden dimensions that will be clear in the context \\ \hline
        $\mathcal{A}=\big\{a_{[1]}, \cdots, a_{[k]}\big\}$ & Action token vocabulary (e.g. \texttt{view}, \texttt{click}, \texttt{like}) \\ \hline
        $y^{(u,j)}_t \in \mathcal{A}$ & Action taken by user $u$ on the $j$-th candidate at serving event $t$ \\ \hline 
        $c_{t}^{(u,j)} \in \mathcal{I}$ & Item identity of the $j$-th candidate served to user $u$ at event $t$ \\ \hline
        $x_t^{(u,j)}$ & Feature vector for $(u, c_{t}^{(u,j)})$ at event $t$, including user feature , item feature, user-item pair feature , and the event's context feature \\ \hline
        $h_{\tilde{t}}^{(u)} \in \mathbb{R}^d$ & Tokenized user behavior at time $\tilde{t}$, where we use $\tilde{t}$ to explicitly suggest that user behavior time is not necessarily aligned with event time $t$  \\ \hline 
        $\mathcal{E}^{(u)}$, $\mathcal{Y}^{(u)}$, $\mathcal{H}^{(u)}$ & Serving event stream, feedback stream, and user behavior stream, defined in Eq.~(\ref{eqn:serving-stream}), (\ref{eqn:feedback-stream}), (\ref{eqn:behavior-stream}) \\ \hline
    \end{tabular}
    }
    \caption{Notations}
    \label{tab:notation}
\end{table}



\begin{figure*}
    \centering
    \includegraphics[width=\textwidth]{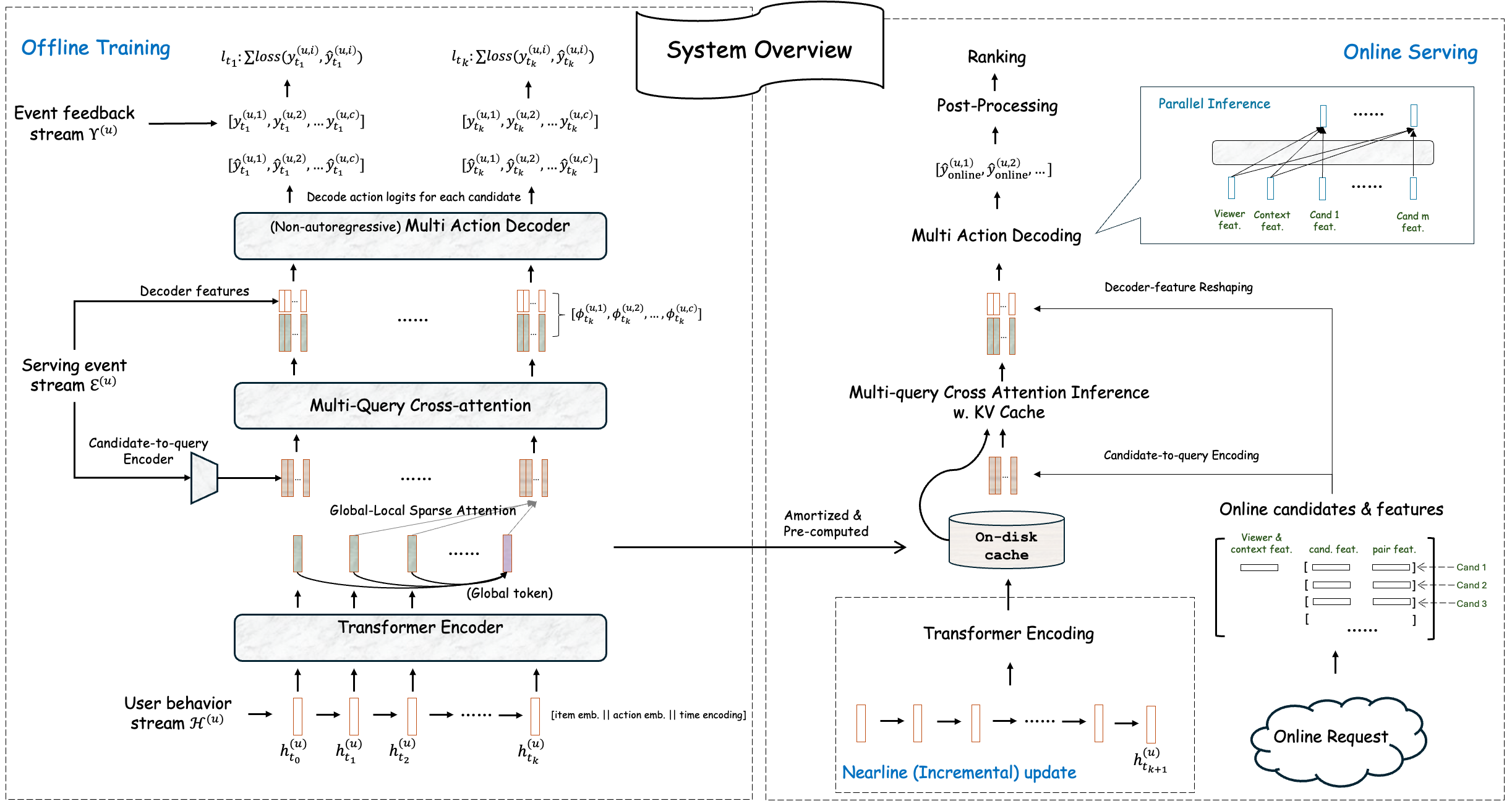}
    \caption{\small End-to-end system overview of TransX, illustrating the unified design across offline training and online serving. \emph{Left}: offline sequence-to-sequence training with a single forward--backward pass per user. \emph{Right}: At online serving time, candidate queries attend to cached user behavior representations via grouped multi-query cross attention, followed by parallel (amortized) non-autoregressive action decoding.
    }
    \label{fig:system}
\end{figure*}

We introduce the notation and terminology for our sequence-to-sequence transduction framework (summarized in Table~\ref{tab:notation}).
Unlike recent frameworks that collapse all user signals into a single sequence, our formulation explicitly distinguishes between \emph{system-driven serving events} and \emph{user-driven behavior streams} (illustrated in the left panel of Figure~\ref{fig:system}). This separation allows us to naturally model the user action token generation process conditioned on both the system’s exposures and the user’s broader behavioral context, without enforcing artificial alignment between system serving event and user behavior.

\paragraph{System Serving Event Stream.}
At each serving event $t$, the recommendation system presents a slate of $m$ candidate items to user $u$. In addition to the user feature and event context feature, each candidate is associated with its item feature and user-item pair feature.
Formally, a serving event is represented as:
\begin{equation}
\label{eqn:serving-stream}
\mathcal{E}_{t}^{(u)} :=
\Big\{ \big(c_{t}^{(u,j)}, x_{t}^{(u,j)} \big) \Big\}_{j=1}^{m_t}.
\end{equation}
The serving event stream for user $u$ is the ordered sequence $\mathcal{E}^{(u)} = \{\mathcal{E}_{t}^{(u)}\}_t$.

\paragraph{Event Feedback Stream.}
Following a serving event, the user may generate explicit or implicit feedback (e.g., \texttt{click}, \texttt{like}, or \texttt{no-action}) for each candidate.
We denote the feedback associated with $\mathcal{E}_{t}^{(u)}$ as:
\begin{equation}
\label{eqn:feedback-stream}
\mathcal{Y}_{t}^{(u)} :=
\big\{ y_{t}^{(u,j)} \big\}_{j=1}^{m},
\end{equation}
where each $y_{t}^{(u,j)} \in \mathcal{A}$.
The feedback stream $\mathcal{Y}^{(u)} = \{\mathcal{Y}_{t}^{(u)}\}_t$ is conditionally generated given the serving stream.

\paragraph{User Behavior Stream.}
In the meantime, users generate a stream of general behaviors that may be unrelated to the targeted recommendation system’s serving events. For example, when modeling an Ads ranking system, a user may still perform actions such as \texttt{search}, \texttt{profile view}, \texttt{connect}, or \texttt{post}.
We represent the tokenized behavior sequence as:
\begin{equation}
\label{eqn:behavior-stream}
\mathcal{H}^{(u)} := \{ h_{\tilde{t}}^{(u)} \},
\end{equation}
where behavior timestamps $\tilde{t}$ are not necessarily aligned with those of $\mathcal{E}^{(u)}$ or $\mathcal{Y}^{(u)}$. This stream provides the historical context for the user's evolving intent.
For brevity, we assume the tokenization of the behavior behavior is given.



\paragraph{The Action Transduction Problem}
Given this multi-stream formulation, we frame the recommendation task as \textbf{sequence-to-sequence action transduction}. Specifically, for a serving event at time $t$, the model must decode the joint action tokens for all candidates, conditioned on the current serving slate and the prefix of the user's historical behavior:
\begin{equation}
\label{eqn:objective}
y_t^{(u,1)}, \ldots, y_t^{(u,m_t)} 
\sim
p\!\left(
\mathcal{Y}_t^{(u)}
\,\middle|\,
\{x_t^{(u,1)}, \ldots, x_t^{(u,m_t)}\}, \mathcal{H}_{<t}^{(u)}
\right),
\end{equation}
By treating recommendation as a transduction task, the model learns to map the "crossing" of the behavior and serving streams into a predictive sequence of user responses.

\section{Methodology}
\label{sec:solution}

In this section, we detail the TransX architecture and its corresponding training and serving system infrastructure. We frame the recommendation task as a sequence-to-sequence transduction problem, mapping an input behavior stream and a serving event stream to a discrete action token stream.

\subsection{TransX Model Architecture}
\label{sec:architecture}

TransX adopts an asymmetric encoder–decoder structure designed to model the crossing of two distinct data processes: the user-driven behavior stream and the system-driven serving stream.

\subsubsection{User Behavior Stream Encoder}

The user behavior stream encoder learns the temporal evolution of a user’s activity history. Unlike monolithic models that directly process raw behavior tokens alongside candidate features, our encoder operates independently, allowing for asynchronous, nearline computation. We denote it by:
\begin{equation*}
\Phi_{\text{behavior}}\!\left(\mathcal{H}^{(u)}_{<t}\right).
\end{equation*}
The encoder consists of Transformer blocks following the standard
dot-product self-attention with causal attention mask, and feed-forward network (FFN) with residual connections and layer normalization \cite{vaswani2017attention}.
Each behavior token is represented by the concatenation of:
(i) a subject embedding (e.g. the embedding of a clicked or viewed item),
(ii) a behavior-type embedding (e.g. \texttt{profile view}, \texttt{search}), and
(iii) a relative position encoding \cite{dai2019transformer}.
This representation allows the encoder to jointly capture \emph{what} action occurred, \emph{on which entity}, and \emph{when} it occurred.

The encoder outputs a sequence of contextualized behavior representations that capture both short-term dynamics and longer-term user intent. However, these representations are not conditioned on the serving event and are therefore insufficient for predicting feedback on specific candidate items. To address this, we fuse the encoded behavior stream with candidate representations to form \emph{candidate-conditioned behavior embeddings} that explicitly model their interaction under the serving context.

\subsubsection{"Stream Crossing" via Grouped Sparse Cross-Attention}
The behavior stream and the serving event stream capture complementary but heterogeneous information: the former summarizes latent user intent from historical activities, while the latter represents the candidate items exposed under the current context. Predicting exposure-conditioned user feedback thus requires an efficient and principled mechanism to align these two streams.

\paragraph{Why cross attention?}
Naively merging the two streams -- e.g. via concatenation or shared self-attention -- fails to respect their asymmetric roles and temporal semantics \cite{qin2020user}.
The behavior stream provides historical context, whereas the serving event stream is instantiated at serving time and defines the prediction targets.
An effective interaction mechanism must therefore allow each candidate to selectively attend to relevant behavioral signals without collapsing the two streams into a single sequence. Cross attention was exactly designed for this purpose \cite{lu2019vilbert}.

\paragraph{Multi-query sparse cross-attention formulation.}

To enable each candidate in a serving event to selectively attend to relevant behavioral signals while maintaining efficiency, TransX adopts a \emph{multi-query cross-attention} design \cite{shazeer2019fast,ainslie2023gqa}.
In this formulation, all candidates within the same serving event share the same
key--value representations derived from the user behavior stream, while maintaining candidate-specific queries.
This design preserves candidate-level specificity while avoiding redundant computation over the behavior stream.
Let $\Phi_{\text{candidate}}(\mathcal{E}_t^{(u)}) \in \mathbb{R}^{m \times d}$ denote the
candidate representations produced by a candidate-to-query encoder
$\Phi_{\text{candidate}}(\cdot)$:
\begin{equation}
\Phi_{\text{candidate}}(\mathcal{E}_t^{(u)}) :=
\Big[\,
\Phi_{\text{candidate}}\!\big(x_{t}^{(u,1)}\big),\ldots,
\Phi_{\text{candidate}}\!\big(x_{t}^{(u,m)}\big)
\Big]^{\top}.
\end{equation}
A simple instantiation of $\Phi_{\text{candidate}}(\cdot)$ is a multi-layer perceptron.

Let $\Phi_{\text{behavior}}(\mathcal{H}^{(u)}_{<t}) \in \mathbb{R}^{L \times d}$ denote the encoded user behavior stream. For each cross-attention head, we compute candidate-specific queries and
\emph{shared} keys and values:
\begin{equation}
\begin{aligned}
&\mathbf{Q} = \Phi_{\text{candidate}}(\mathcal{E}_t^{(u)}) \mathbf{W}_Q, \\
&\mathbf{K} = \Phi_{\text{behavior}}(\mathcal{H}^{(u)}_{<t}) \mathbf{W}_K, \mathbf{V} = \Phi_{\text{behavior}}(\mathcal{H}^{(u)}_{<t}) \mathbf{W}_V,
\end{aligned}
\end{equation}
where $\mathbf{K}$ and $\mathbf{V}$ are shared across all $m$ candidate queries in the serving event.

\paragraph{Local--global sparse cross attention.}
Directly attending to all past USER behavior tokens is computationally expensive and unnecessary, as user intent is often expressed through a combination of
\emph{recent local behaviors} and \emph{long-range global preferences}.
We therefore introduce a local--global sparse attention pattern that restricts
cross attention to a structured subset of the behavior stream and a global anchor.

Specifically, for each candidate query, we define a sparse index set
$\mathcal{I}_{t} = \mathcal{I}_{t,\text{local}} \cup \mathcal{I}_{t,\text{global}}$, where:
\begin{itemize}[leftmargin=*]
    \item $\mathcal{I}_{t, \text{local}}$ contains the most recent $L_{\text{local}}$
    behavior tokens preceding event $t$, capturing short-term intent;
    \item $\mathcal{I}_{t, \text{global}}$ contains an globally-pooled anchor token -- such as the attention-pooled global representation up to time $t$ -- capturing long-term preferences.
\end{itemize}

Let $\mathbf{K}_{\mathcal{I}}$ and $\mathbf{V}_{\mathcal{I}}$ denote the keys and values restricted to a sparse index $\mathcal{I}$.
The sparse cross attention is then computed as: 
\begin{equation}
\label{eqn:sparse-cross-attn}
\text{softmax}\!\left(
\frac{\mathbf{Q}\mathbf{K}_{\mathcal{I}}^{\top}}{\sqrt{d_h}}
\right)\mathbf{V}_{\mathcal{I}}.
\end{equation}

We then employ multi-head attention with the above sparse attention setup, followed by residual connections and feed-forward network (FFN). With the shorthand for candidate-to-query encoding and behavior stream encoding: $\Phi_{c,t}:=\Phi_{\text{candidate}}\left(\mathcal{E}_t^{(u)}\right)$ and $\Phi_{b,t}:=\Phi_{\text{behavior}}\left(\mathcal{H}^{(u)}_{<t}\right)$, the full cross attention in TransX is: 
\begin{equation}
\begin{aligned}
\text{crossAttend}\!\left(\mathcal{E}_t^{(u)}, \mathcal{H}^{(u)}_{<t}\right)
&= \text{LN}\Big(
    \Phi_{c,t}
    + \text{FFN}\big( \tilde{\Phi}_{c,t} \big)
\Big), \\[4pt]
\tilde{\Phi}_{c,t}
&= \text{LN}\Big(
    \Phi_{c,t}
    + \text{MHA}_{\text{LG}}\!\left(\Phi_{c,t}, \Phi_{b,t}\right)
\Big).
\end{aligned}
\end{equation}
where $\text{LN}(\cdot)$ denotes layer normalization, $\text{MHA}_{\text{LG}}(\cdot,\cdot)$ denotes multi-head attention with global-local sparse attention pattern.

\begin{remarkbox}
\textbf{Remark.}
The sparse cross-attention design yields two practical benefits for online serving:
\begin{itemize}[leftmargin=*]
    \item the online computation cost scales as
    $\mathcal{O}\!\big(mL_{\text{local}}\big)$ per event rather than
    $\mathcal{O}(mL)$, due to sparse attention and shared keys/values;
    \item in online serving, $\Phi_{\text{behavior}}(\mathcal{H}^{(u)}_{<t})$ and the corresponding keys and values can be cached and incrementally updated, further
    improving serving efficiency at scale.
\end{itemize}
\end{remarkbox}


\subsubsection{Action Transduction Decoder}

The final stage of the architecture is the Parallel Action Decoder, which maps the crossed stream representations and serving event features $x_t^{(u,i)}$ into the action token space.
Unlike conventional recommendation models that score candidates independently,
TransX formulates action prediction as a \emph{parallel, grouped decoding} problem
over a discrete action vocabulary:
\begin{equation*}
p\!\left(
\mathcal{Y}_t^{(u)}
\,\middle|\,
\{x_t^{(u,1)}, \ldots, x_t^{(u,m_t)}\}, \mathcal{H}_{<t}^{(u)}
\right)=\prod_{i=1}^{m_t} p\!\left(y_{y}^{(u,i)} \,\middle|\, x_t^{(u,i)} , \mathcal{H}_{<t}^{(u)}\right) ,
\end{equation*}

\paragraph{Decoder formulation.}
For a serving event $t$ and user $u$, let
\begin{equation*}
    \mathbf{Z}_{t}^{(u)} := \text{crossAttend}\left(\mathcal{E}_t^{(u)}, \mathcal{H}^{(u)}_{<t} \right) \in \mathbb{R}^{m \times d}
\end{equation*}
denote the output of the cross-attention block, where each row corresponds to a
candidate-conditioned representation for one candidate item.
The goal of the decoder is to predict, for each candidate $c$, a distribution over
action tokens in $\mathcal{A}$ given $\mathbf{Z}_{t}^{(u)}$ and all the event features $\mathcal{X}(\mathcal{E}_t^{(u)})$.

To this end, we adopt a query--key softmax decoder. Specifically, we maintain a set of learnable action token embeddings
\begin{equation}
\mathbf{Q}_{\text{dec}} =
\big[
\mathbf{q}_{\text{dec}}^{(a_1)}, \ldots, \mathbf{q}_{\text{dec}}^{(a_k)}
\big] \in \mathbb{R}^{k \times d},
\end{equation}
where each $\mathbf{q}_{\text{dec}}^{(a)}$ corresponds to an action token
(e.g. \texttt{click}, \texttt{no-click}, \texttt{connect}).

For each candidate $c$ in the serving event, we map its candidate-conditioned
representation $\mathbf{z}_{t}^{(u,c)}$ and event features $x_t^{(u,c)}$ to a decoder key via a non-linear feature-crossing head $f_{\text{dec}}(\cdot)$:
\begin{equation}
\mathbf{k}_{\text{dec}}^{(u,c)} =
f_{\text{dec}}\!\left(\mathbf{z}_{t}^{(u,c)}, x_{t}^{(u,c)}\right),
\end{equation}
where the detailed formulation of $f_{\text{dec}}(\cdot)$ is discussed in Appendix \ref{sec:actionhead-append}, as it is not the primary focus of this work. Here, we empathize that $f_{\text{dec}}(\cdot)$ is explicitly designed to support parallel computation for each candidate, as illustrated in the top right corner of Figure \ref{fig:system}.

The action distribution for candidate $c$ is then computed as:
\begin{equation}
\label{eqn:action-decoding}
p\!\left(y_{t}^{(u,c)} \mid \mathcal{E}_{t}^{(u)}, \mathcal{H}_{<t}^{(u)}\right)
=
\text{softmax}\!\left(
\mathbf{Q}_{\text{dec}} \, \mathbf{k}_{\text{dec}}^{(u,c)}
\right).
\end{equation}

The above query--key softmax decoder setup is primarily motivated by \textbf{calibration considerations}, which are critical in large-scale industrial recommenders \cite{steck2018calibrated,he2014practical}.
By decoupling the action vocabulary from candidate and behavior representations,
the decoder isolates action probability estimation from upstream representation learning, preventing uncontrolled logit scaling and cross-task interference.
This separation enforces a single, shared normalization over actions, yielding
probabilities that remain comparable and stable across candidates and serving events.

\paragraph{Parallel grouped decoding.}
Importantly, the decoder is applied \emph{in parallel} to all $m$ candidates within the same serving event, sharing the same action token queries
$\mathbf{Q}_{\text{dec}}$.
This design naturally supports the multi-action-token formulation introduced in
Section~\ref{sec:prelim}, while preserving candidate-specific conditioning
from the multi-query cross attention step.


\subsection{Streaming Seq-to-Seq Training}
\label{sec:training}

TransX is trained using a \textbf{single-pass, sequence-to-sequence paradigm} that aligns naturally with the streaming data of industrial RecSys. Unlike the "per-event" training used in traditional DLRMs -- which repeatedly re-processes user history for every label -- our transduction formulation maps the entire sequence of serving events and compute the loss and gradient in one forward-backward pass.


\paragraph{Sequence-to-sequence formulation.}
For a given user $u$, let $\{\mathcal{E}_t^{(u)}\}_{t=1}^{T_u}$ denote the sequence of serving events and $\{\mathcal{Y}_t^{(u)}\}_{t=1}^{T_u}$ the corresponding event-level feedback.
At each event $t$, TransX predicts $\mathcal{Y}_t^{(u)}$ conditioned on the serving event $\mathcal{E}_t^{(u)}$ and the prefix of the behavior stream
$\mathcal{H}_{<t}^{(u)}$, yielding a grouped, non-autoregressive seq-to-seq mapping:
\begin{equation}
\label{eqn:streaming-seq2seq}
\big( \mathcal{H}^{(u)}, \{\mathcal{E}_t^{(u)}\}_{t=1}^{T_u} \big)
\;\xrightarrow{\;\;\text{TransX}\;\;}
\{\mathcal{Y}_t^{(u)}\}_{t=1}^{T_u}.
\end{equation}
Each serving event acts as a decoding step, while avoiding autoregressive dependencies across events.

\paragraph{Single-pass forward and backward computation.}
The training loss for user $u$ aggregates contributions from all serving events:
\begin{equation}
\label{eqn:seq2seq-loss}
\mathcal{L}^{(u)} =
\sum_{t=1}^{T_u}
\sum_{c=1}^{m_t}
\ell\!\left( y_t^{(u,c)}, \hat{y}_t^{(u,c)} \right),
\end{equation}
where $\hat{y}_t^{(u,c)}$ is produced by the parallel multi-action decoder.
This event-factorized objective allows losses from the entire sequence to be accumulated in a single forward--backward pass, without event-level unrolling.

\paragraph{Streaming behavior encoding and amortization.}
During training, the behavior encoder maintains an incremental representation
$\Phi_{\text{behavior}}(\mathcal{H}_{<t}^{(u)})$ as new activity tokens arrive.
Event-level modules—including multi-query local-global sparse cross-attention and parallel multi-action decoding—are invoked only at serving events.
This decoupling amortizes the cost of sequence encoding across events, enabling TransX to scale to long histories while keeping per-event computation lightweight.

\paragraph{Causal attention and prefix reuse.}
When the behavior stream encoder employs \emph{causal self-attention}, each behavior representation depends only on past activities.
As a result, the entire behavior stream $\mathcal{H}^{(u)}$ can be encoded once in a single left-to-right forward pass, producing prefix-consistent representations that are reused across all serving events.
For any event at time $t$, TransX simply slices the precomputed prefix $\Phi_{\text{behavior}}(\mathcal{H}^{(u)}_{<t})$ without re-encoding.
This property is a key enabler of TransX’s single-pass seq-to-seq training and further enables the nearline incremental updates of $\Phi_{\text{behavior}}(\mathcal{H}^{(u)}_{<t})$ used during online serving.

\paragraph{Customized attention masks.}
TransX applies attention masks tailored to the semantics of each module.
The behavior encoder uses a causal mask to prevent information leakage from future
activities, while cross attention employs sparse, event-level masks (e.g., local--global masks) to restrict attention to the most relevant subsets of historical behaviors. These customized masks enforce correct temporal dependencies while reducing unnecessary computation.



\subsection{Online Serving Optimizations}
\label{sec:inference}

\begin{figure}[htb]
    \centering
    \includegraphics[width=\linewidth]{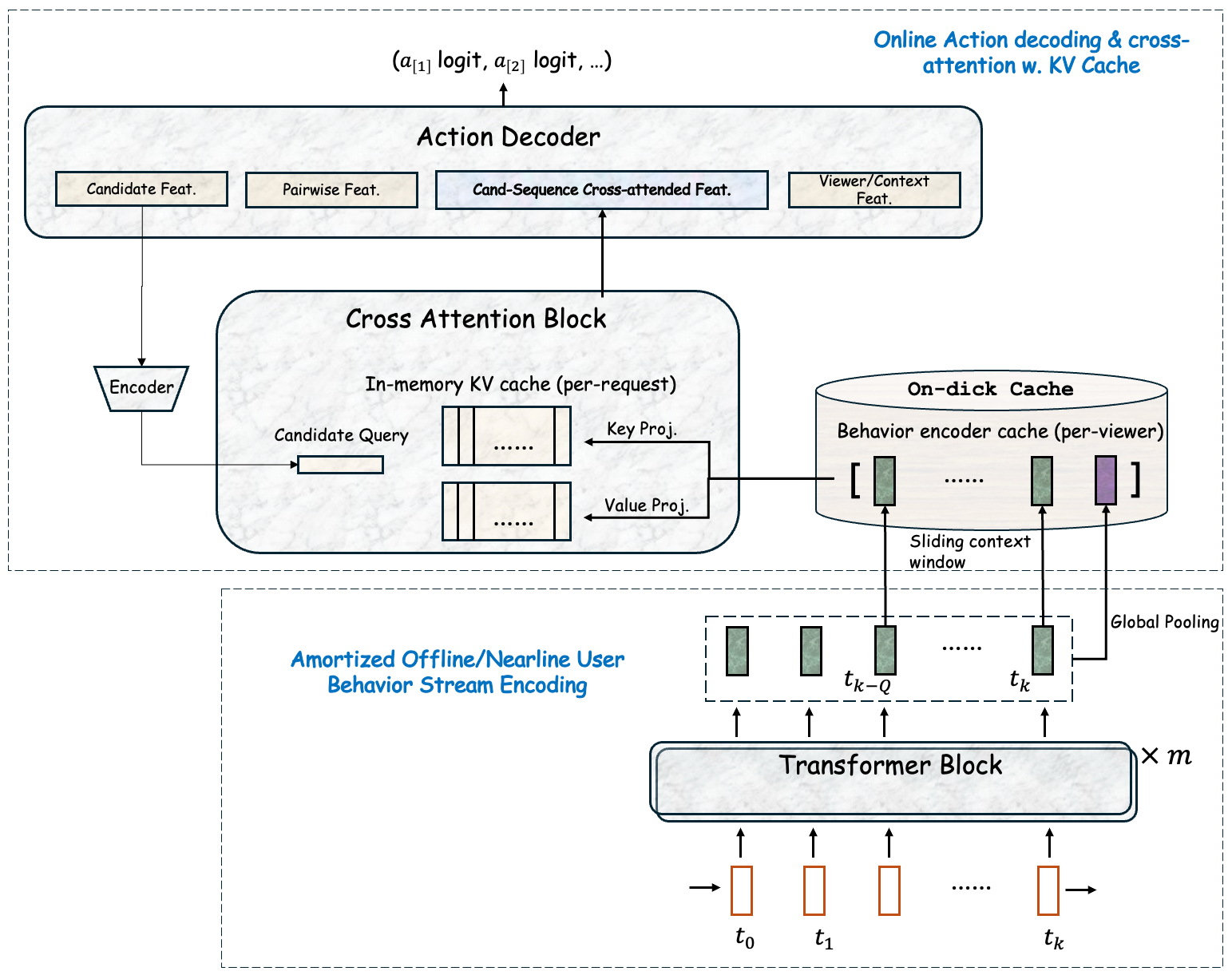}
    \caption{\small Overview of TransX’s online inference pipeline with model--infra co-design. User behavior streams are encoded offline/nearline, cached per user, and incrementally updated nearline.
    At serving time, only lightweight event-level computation is performed:
    candidate queries attend to a cached, local--global subset of behavior representations via grouped multi-query cross attention with shared key--value caches, followed by a parallel non-autoregressive action decoder.
    }
    \label{fig:inference}
\end{figure}

The scalability of TransX is realized through a tight coupling of the model architecture and the production serving pipeline (Figure~\ref{fig:inference}) to operate under strict online latency and cost constraints,
where recommendations must be produced within hundred milliseconds while serving large candidate sets (i.e. $\sim 10^4$) at hundred of thousands of requests per second.

\paragraph{Amortized nearline encoding and caching}
The viewer behavior sequence encoder learns rich contextual representations for user activities but is also the most computationally expensive module.
By amortizing this computation and moving it off the critical online path, TransX shifts the bulk of model cost away from the online serving path. User behavior representations are computed (refreshed) nearline as activity tokens arrive and are stored in a per-viewer on-disk cache.
This design not only reduces online CPU/GPU usage, but also enables the model to scale to ultra-long behavior sequences and more expressive tokenization schemes that would otherwise be infeasible under strict latency budgets. To further reduce storage and memory overhead, we apply post-training quantization to the cached representations (i.e. use \texttt{bf16} instead of \texttt{float32}) \cite{dao2022flashattention} and observed negligible impact on model performance.

\paragraph{Global--local sparse attention.}
The global--local sparse attention pattern in cross attention to further reduces both computation and cache I/O. Rather than attending to the full behavior history, candidate queries attend only to a small set of recent behavior tokens within a sliding window and a globally pooled summary representation.
This sparsity pattern significantly lowers the cost of key--value retrieval and attention computation.

\paragraph{Grouped multi-query cross attention with KV caching.}
During online serving, only the lightweight cross-attention and decoding are performed. By caching the behavior-derived Keys and Values at the start of a request -- which is commonly referred to as KV caching \cite{tay2020efficient} -- we ensure that the marginal cost of scoring additional candidates is nearly constant, scaling linearly with the number of candidates but remaining independent of the total behavior sequence length.


\subsection{Complexity Analysis}
\label{sec:complexity}

Let $L_{\text{local}} \ll L$ be the size of the local attention window and $T$ be the number of serving events. For simplicity, assume the decoding head $f_{\text{dec}}(\cdot)$ is a linear function. We provide the detailed derivations in Appendix \ref{sec:complexity-append} and summarize the analysis results below.

Compared to the typical sequential and GR models, by enabling amortized and cachable sequence encoding, TransX approximately reduces training complexity from $\mathcal{O}(T L^2 d)$ to $\mathcal{O}(L^2 d + T)$, up to constant factors. As a result, \emph{the number of training events contributes only additively to the overall training cost, rather than multiplicatively as in the DLRM paradigm}.
At serving time, \emph{TransX avoids any dependence on the full behavior length $L$},
achieving $L_{\text{local}}$-bounded online computation at the cost of a constant
$\mathcal{O}(L_{\text{local}} d)$ memory footprint. Together, they enable the deployment of our advanced Transformer-based action transduction model under sub-100ms latency SLAs without incurring significant online hardware cost.
Empirically, we observe the seq-to-seq training formulation reduces per-epoch training time by  $\sim 50\%$ relative to the per-event training paradigm prevalent in the DLRM era; and the serving optimizations can reduce online computation by $\sim 80\%$, enabling deployment of an advanced encoder-decoder model without incurring prohibitive hardware cost.

\section{Offline and Online Experiments}
\label{sec:experiment}

\subsection{Offline Model Benchmarking}
\label{sec:offline}

\paragraph{Baseline.} 
We compare TransX with the following baselines on our production data:
\begin{itemize}[leftmargin=*]
    \item \textbf{LiDLRM} -- LinkedIn's personalized deep learning recommendation model currently used in production \cite{borisyuk2024lirank};
    \item \textbf{SASRec} -- the canonical Transformer-based sequential recommender
    that models user behavior as sequence and predicts the next item via self-attention \cite{kang2018self};
    \item \textbf{TransAct} -- Pinterest's Transformer-based user action model that encodes users’ recent activity sequences to extract short-term preferences for next-item CTR prediction \cite{xia2023transact};
    \item \textbf{GRM} -- a generative recommendation model built similar to Meta's GRM \cite{zhai2024actions} to represent the class of SOTA model architecture that collapsees all (heterogeneous) data into a monolithic token stream for CTR prediction as next-action generation.
\end{itemize}

\paragraph{Recommendation task.}
Since most of the baselines considered in this study are designed for click-through rate (CTR) prediction, we restrict the prediction target to the \texttt{click} action to enable fair and consistent comparisons.
This reduces the learning objective to a binary classification problem with label
$y_t^{(u,c)} \in \{0,1\}$, indicating whether user $u$ clicks candidate $c$ at serving event $t$.
For offline evaluation, we report AUC, AUPR, and group-averaged AUC (gAUC).
Here, gAUC is computed by first evaluating AUC independently for each user and then averaging across users, assigning equal weight to each user regardless of activity level.

\paragraph{Production Dataset.}
We conduct experiments on real-world user behavior, system serving event, and event
feedback data collected from \textbf{LinkedIn’s largest social recommendation
application}, which collectively serve hundreds of millions of daily user interactions.
Leveraging TransX’s ability to scale to long user histories, we construct
$\{\mathcal{H}^{(u)}_t\}_t$, $\{\mathcal{E}^{(u)}_t\}_t$, and $\{\mathcal{Y}^{(u)}_t\}_t$ using a 180-day look-back window.
All datasets are split chronologically into training, validation, and test sets to
respect temporal causality and prevent information leakage. Since the positive action token rate is $\sim 3\%$, for each serving event $\mathcal{E}^{(u)}_t$ in training dataset, we apply downsampling to "\texttt{no action}" token so that the number of negative candidates matches the number of positive candidates. For validation and testing, we do not apply downsampling so the offline metrics are computed on the actual distribution.

\paragraph{Model configrations}
For TransX, the behavior sequence encoder $\Phi_{\text{behavior}}\!\left(\mathcal{H}^{(u)}_{<t}\right)$ uses two standard causal Transformer layers with relative position encoding \cite{dai2019transformer}. For cross attention, we set $L_{\text{local}}=10$ and use attention pooling to generate the global token for each event timestamp. For other baseline models, while keeping their original architecture as much as possible and aligning relevant configurations with TransX (e.g. use the same causal mask and local attention setup when applicable) on our production data. We also modify the hidden dimensions so all models have a similar offline compute budget measured by \textbf{MFLOPs per candidate item scoring} offline. More details can be found in Appendix~\ref{sec:benchmark-append}.

\paragraph{Offline results}
The key observations from Table~\ref{tab:benchmark} are:

\begin{itemize}[leftmargin=*]
    \item \textbf{State-of-the-art accuracy with dramatically lower serving cost.}
    Under a comparable compute budget (per candidate scoring), TransX matches or exceeds the strongest industrial sequential and GR baselines (e.g., TransAct \cite{xia2023transact} and MTGR \cite{han2025mtgr}) across all offline metrics (AUC, AUPR, and gAUC), while requiring only $\sim20\%$ of their online compute.
    This large gap between offline and online cost highlights TransX’s superior production scalability and substantially more favorable serving economics.

    \item \textbf{Clear architectural advantages over legacy sequential models.}
    Compared with baselines based on earlier architectures (e.g. scaled up LiDLRM \cite{borisyuk2024lirank} and SASRec \cite{kang2018self}), TransX achieves significant and consistent gains at similar or lower compute budgets. These improvements cannot be attributed to model size alone and instead demonstrate the effectiveness of TransX’s architectural design for modern large-scale sequential recommendation.
\end{itemize}

\begin{table}[hbt]
    \centering
    \begin{tabular}{c|c|c|c|p{1.8cm}}
    \hline
         & AUC  & AUPR  & gAUC & MFLOPs (offline/online) \\ \hline \hline
        $\text{LiDLRM}_{\text{prod}}$ & 0.846  & 0.257   & 0.636   & $0.8/0.8\times 10^2$  \\
        $\text{LiDLRM}_{\text{large}}$ & 0.852  & 0.264  & 0.641  & $3.9/3.9\times 10^2$  \\
        SASRec & 0.851  &  0.260  & 0.641   & $3.5/3.5\times 10^2$  \\
        $\text{TransAct}_{\text{concat}}$ & 0.853  & 0.262  & 0.629  & $3.6/3.6\times 10^2$ \\
        $\text{TransAct}_{\text{append}}$ & 0.860  & 0.271  & 0.644   & $4.0/4.0\times 10^2$ \\
        GRM & 0.858  & 0.269  &  0.645  & $4.1/4.1\times 10^2$ \\ \hline \hline
        \textbf{TransX} & \textbf{0.862}  & \textbf{0.273}  & \textbf{0.648}   & $\textbf{3.8/0.6} \times 10^2 $ \\ 
        $\text{TransX}_{\text{-crossAttn}}$ & $-13.4\%$  & $-10.6\%$  & $-7.3\%$  &  $3.6/0.4\times 10^2$ \\
        $\text{TransX}_{\text{-encoder}}$ & $-28.3\%$ & $-22.5\%$  & $-15.8\%$ &   $0.7/0.7\times 10^2$  \\
        $\text{TransX}_{\text{-sparseAttn}}$ & $+4.5\%$  & $+3.6\%$  & $+2.1\%$  &  $5.1/2.2\times 10^2$ \\ \hline \hline
    \end{tabular}
    \caption{\small Offline benchmarking results. $\text{LiDLRM}_{\text{prod}}$ is the current production model, $\text{LiDLRM}_{\text{large}}$ is the scaled-up version to match the MFLOPs of TransX. $\text{TransAct}_{\text{concat}}$ and $\text{TransAct}_{\text{append}}$ are the two variants of TransAct model as introduced in \cite{xia2023transact}. For the ablated TransX variants reported in the lower panel, metrics are presented as relative changes for clearer comparison. The results are averaged over 5 independent runs.
    }
    \label{tab:benchmark}
\end{table}
\vspace{-0.3cm}

\subsection{Ablation Studies}
\label{sec:ablation}

We first ablate the key architectural components of TransX to quantify their individual contributions:
\begin{itemize}[leftmargin=*]
    \item $\text{TransX}_{\text{-crossAttn}}$, which removes the cross-attention module and instead directly appends the most recent $L_{t,\text{local}}$ behavior encodings to the event-level user feature $x_t^{(u)}$;
    \item $\text{TransX}_{\text{-encoder}}$, which removes the encoder stack and directly applies cross-attention between the behavioral sequence and event candidates;
    \item $\text{TransX}_{\text{-sparseAttn}}$, in which the cross-attention module uses full global attention rather than the proposed sparse attention.
\end{itemize}
Table \ref{tab:benchmark} suggests that removing the cross-attention module leads to consistent and substantial degradation across all offline metrics, indicating that explicit interaction modeling between behavioral history and serving events is essential.
Eliminating the encoder stack causes the most severe performance drop, confirming the necessity of learnt sequence encoding for capturing behavioral interaction and dependencies.
Finally, replacing the proposed sparse attention with full global attention yields only marginal accuracy gains while incurring significantly higher offline and online computation, highlighting the favorable accuracy–efficiency trade-off achieved by sparse attention.
Together, these results validate that TransX’s architectural design choices are both necessary for strong predictive performance and crucial for maintaining production-grade computational efficiency.

The second set of ablation studies focuses on the encoder configuration, including the number of Transformer layers, position bias and attention type, and the sparse cross attention design, including the local context window size $L_{\text{local}}$ and global pooling method. 
The ablation results in Tables~\ref{tab:encoder-ablation} and~\ref{tab:cross-attn-ablation} highlight the effectiveness of using a causal attention mask together with relative position encoding in the behavior encoder, achieving competitive performance while retaining favorable computational efficiency.
For sparse cross-attention, combining a learnable global token with attention pooling and small-to-medium local context windows yields favorable accuracy–compute trade-off, delivering strong performance gains while controlling computational cost.

\subsection{Production Deployment and A/B Testings}
\label{sec:production}

Productionizing TransX to hundreds of thousands of QPS under a strict p99 latency SLA on the order of hundreds of milliseconds is highly challenging.
As shown in Table~\ref{tab:latency-reduction}, our model–infrastructure co-design enables user-level amortized behavior sequence encoding with caching, which, when further combined with request-level KV caching, substantially reduces p99 latency.
In particular, under realistic serving conditions where a viewer on average scores approximately 500 candidates per request, these optimizations reduce p99 end-to-end latency by up to 80\%, leading to a hardware cost that is comparable to the baseline DLRM model currently in production \cite{borisyuk2024lirank}. Due to page limit, we provide additional deployment details in Appendix~\ref{sec:coldstart-append}, covering cold-start handling, cache management, and nearline updates with a composition of event-based and time-based triggers. Together, these mechanisms ensure robustness, scalability, and stable online performance under real-world traffic conditions.

\begin{table}[htb]
\centering
    \begin{tabular}{p{1.3cm}|ccc|ccc}
    \hline
     & \multicolumn{3}{c|}{Amortized Seq. Encoding} & \multicolumn{3}{c}{+ KV Caching} \\ 
    & \multicolumn{3}{c|}{and Caching} & \multicolumn{3}{c}{} \\
    \#Candidate & n=300 & n=400 & n=500 & n=300 & n=400 & n=500 \\ \hline
    Latency reduction & -58\% & -64\% & -73\% &  -63\%  &  -71\%  & -83\% \\ \hline
    \end{tabular}
\caption{\small P99 end-to-end serving latency reduction achieved by TransX’s serving optimizations, benchmared on real-world traffic pattern with an average 80K request per second under varying candidate sizes $n$.}
\label{tab:latency-reduction}
\end{table}

As shown in Table~\ref{tab:ab-results}, TransX delivers substantial and online gains when deployed for A/B test on LinkedIn’s largest social recommendation task that serves billions of users daily. Relative to the DLRM baseline, TransX improves click-through rate by 6.0\% and conversion rate by 4.4\% (with a platform-level impact of 0.26\% increase in daily active users and a 0.08\% increase in home feed views), representing the largest lifts observed in recent years from this application. From a cost perspective, the MFLOPs results in Table~\ref{tab:benchmark} demonstrate that TransX can scale total computation by 5–6$\times$ without incurring additional online compute cost, which matches our actual cost saving in practice, leading to the following remark.

\begin{table}[ht]
    \centering
    \begin{tabular}{cccc}
    \hline \hline
    \textbf{Task Click-} & \textbf{Task Conver} & \textbf{Daily Active} & \textbf{Home Feeds} \\  
    \textbf{through-rate} & \textbf{-sion rate} & \textbf{User} & \textbf{View} \\ \hline
    \textcolor{green!60!black}{+6.0\% $\uparrow$} &
    \textcolor{green!60!black}{+4.4\% $\uparrow$} &
    \textcolor{green!60!black}{+0.26\% $\uparrow$} &
    \textcolor{green!60!black}{+0.08\% $\uparrow$} \\ 
    $(\text{p}<10^{-4})$ & $(\text{p}<10^{-4})$ & $(\text{p}<10^{-4})$ & $(\text{p}<10^{-2})$ \\ \hline \hline
    \end{tabular}
    \caption{\small Online A/B testing results from a 50\%–50\% traffic split comparing TransX against the DLRM baseline on LinkedIn’s main social recommendation task, showing statistically significant improvements across key business metrics. In particular, the observed lifts in CTR and Conversion are the largest achieved in recent years.}
    \label{tab:ab-results}
\end{table}


\begin{remarkbox}
\textbf{Remark: model-infra co-design advancing the Pareto Frontier of Industrial RecSys}. Industrial RecSys is defined by the trade-off between modeling performance and serving cost. TransX represents an example where model-infra co-design can advances the Pareto frontier on both fronts, where "scaling up" recommendation models to drive performance does not necessarily require a 1-to-1 increase in online hardware investment to serve the models.
\end{remarkbox}


\section{Discussion}
\label{sec:discussion}

Beyond the substantial online metric improvements, the success of TransX suggests a promising direction for architecting large-scale industrial recommenders. A core takeaway from this work is that by explicitly modeling the "crossing" of the user behavior stream and the system serving stream, we allow the selective application of expensive Transformer layers where they are most needed, rather than uniformly across a flattened monolithic token stream. 


One promising direction is to extend TransX into a unified retrieval-and-ranking framework that jointly handles candidate generation and action transduction. The sequence–candidate crossing architecture naturally supports this extension by introducing an auxiliary contrastive objective (e.g. InfoNCE) between the user behavior encoding $\Phi_{\text{behavior}}(\cdot)$  and candidate encoding $\Phi_{\text{candidate}}(\cdot)$. This enables TransX to learn retrieval-oriented representations while preserving its action-transduction capability. Because both objectives share the same behavior encoder, the resulting framework remains compatible with the sequence-to-sequence training and amortized serving paradigm, offering a path toward end-to-end recommendation with a single model.

Finally, the transition from discriminative DLRMs to transductive Transformers marks a new era for industrial recommendation. TransX demonstrates that by applying the appropriate structural inductive bias, we can achieve record-level online gains while maintaining the stringent latency and cost profiles. As behavioral histories continue to grow in complexity and duration, the "Stream Crossing" paradigm provides a scalable, principled framework for the next generation of industrial recommender systems.

\bibliographystyle{ACM-Reference-Format}
\balance
\bibliography{references}

@inproceedings{borisyuk2024lirank,
  title={LiRank: Industrial Large Scale Ranking Models at LinkedIn},
  author={Borisyuk, Fedor and Zhou, Mingzhou and Song, Qingquan and Zhu, Siyu and Tiwana, Birjodh and Parameswaran, Ganesh and Dangi, Siddharth and Hertel, Lars and Xiao, Qiang Charles and Hou, Xiaochen and others},
  booktitle={Proceedings of the 30th ACM SIGKDD Conference on Knowledge Discovery and Data Mining},
  pages={4804--4815},
  year={2024}
}

@inproceedings{kang2018self,
  title={Self-attentive sequential recommendation},
  author={Kang, Wang-Cheng and McAuley, Julian},
  booktitle={2018 IEEE international conference on data mining (ICDM)},
  pages={197--206},
  year={2018},
  organization={IEEE}
}

@inproceedings{xia2023transact,
  title={Transact: Transformer-based realtime user action model for recommendation at pinterest},
  author={Xia, Xue and Eksombatchai, Pong and Pancha, Nikil and Badani, Dhruvil Deven and Wang, Po-Wei and Gu, Neng and Joshi, Saurabh Vishwas and Farahpour, Nazanin and Zhang, Zhiyuan and Zhai, Andrew},
  booktitle={Proceedings of the 29th ACM SIGKDD Conference on Knowledge Discovery and Data Mining},
  pages={5249--5259},
  year={2023}
}

@inproceedings{devlin2019bert,
  title={Bert: Pre-training of deep bidirectional transformers for language understanding},
  author={Devlin, Jacob and Chang, Ming-Wei and Lee, Kenton and Toutanova, Kristina},
  booktitle={Proceedings of the 2019 conference of the North American chapter of the association for computational linguistics: human language technologies, volume 1 (long and short papers)},
  pages={4171--4186},
  year={2019}
}

@inproceedings{dai2019transformer,
  title={Transformer-xl: Attentive language models beyond a fixed-length context},
  author={Dai, Zihang and Yang, Zhilin and Yang, Yiming and Carbonell, Jaime G and Le, Quoc and Salakhutdinov, Ruslan},
  booktitle={Proceedings of the 57th annual meeting of the association for computational linguistics},
  pages={2978--2988},
  year={2019}
}

@inproceedings{cheng2016wide,
  title={Wide \& deep learning for recommender systems},
  author={Cheng, Heng-Tze and Koc, Levent and Harmsen, Jeremiah and Shaked, Tal and Chandra, Tushar and Aradhye, Hrishi and Anderson, Glen and Corrado, Greg and Chai, Wei and Ispir, Mustafa and others},
  booktitle={Proceedings of the 1st workshop on deep learning for recommender systems},
  pages={7--10},
  year={2016}
}

@article{guo2017deepfm,
  title={DeepFM: a factorization-machine based neural network for CTR prediction},
  author={Guo, Huifeng and Tang, Ruiming and Ye, Yunming and Li, Zhenguo and He, Xiuqiang},
  journal={arXiv preprint arXiv:1703.04247},
  year={2017}
}

@article{naumov2019deep,
  title={Deep learning recommendation model for personalization and recommendation systems},
  author={Naumov, Maxim and Mudigere, Dheevatsa and Shi, Hao-Jun Michael and Huang, Jianyu and Sundaraman, Narayanan and Park, Jongsoo and Wang, Xiaodong and Gupta, Udit and Wu, Carole-Jean and Azzolini, Alisson G and others},
  journal={arXiv preprint arXiv:1906.00091},
  year={2019}
}

@incollection{wang2017deep,
  title={Deep \& cross network for ad click predictions},
  author={Wang, Ruoxi and Fu, Bin and Fu, Gang and Wang, Mingliang},
  booktitle={Proceedings of the ADKDD'17},
  pages={1--7},
  year={2017}
}

@inproceedings{wang2021dcn,
  title={Dcn v2: Improved deep \& cross network and practical lessons for web-scale learning to rank systems},
  author={Wang, Ruoxi and Shivanna, Rakesh and Cheng, Derek and Jain, Sagar and Lin, Dong and Hong, Lichan and Chi, Ed},
  booktitle={Proceedings of the web conference 2021},
  pages={1785--1797},
  year={2021}
}

@inproceedings{lian2018xdeepfm,
  title={xdeepfm: Combining explicit and implicit feature interactions for recommender systems},
  author={Lian, Jianxun and Zhou, Xiaohuan and Zhang, Fuzheng and Chen, Zhongxia and Xie, Xing and Sun, Guangzhong},
  booktitle={Proceedings of the 24th ACM SIGKDD international conference on knowledge discovery \& data mining},
  pages={1754--1763},
  year={2018}
}

@article{vaswani2017attention,
  title={Attention is all you need},
  author={Vaswani, Ashish and Shazeer, Noam and Parmar, Niki and Uszkoreit, Jakob and Jones, Llion and Gomez, Aidan N and Kaiser, {\L}ukasz and Polosukhin, Illia},
  journal={Advances in neural information processing systems},
  volume={30},
  year={2017}
}

@inproceedings{sun2019bert4rec,
  title={BERT4Rec: Sequential recommendation with bidirectional encoder representations from transformer},
  author={Sun, Fei and Liu, Jun and Wu, Jian and Pei, Changhua and Lin, Xiao and Ou, Wenwu and Jiang, Peng},
  booktitle={Proceedings of the 28th ACM international conference on information and knowledge management},
  pages={1441--1450},
  year={2019}
}

@inproceedings{chen2019behavior,
  title={Behavior sequence transformer for e-commerce recommendation in alibaba},
  author={Chen, Qiwei and Zhao, Huan and Li, Wei and Huang, Pipei and Ou, Wenwu},
  booktitle={Proceedings of the 1st international workshop on deep learning practice for high-dimensional sparse data},
  pages={1--4},
  year={2019}
}

@article{gui2023hiformer,
  title={Hiformer: Heterogeneous feature interactions learning with transformers for recommender systems},
  author={Gui, Huan and Wang, Ruoxi and Yin, Ke and Jin, Long and Kula, Maciej and Xu, Taibai and Hong, Lichan and Chi, Ed H},
  journal={arXiv preprint arXiv:2311.05884},
  year={2023}
}

@inproceedings{zeng2025interformer,
  title={InterFormer: Effective Heterogeneous Interaction Learning for Click-Through Rate Prediction},
  author={Zeng, Zhichen and Liu, Xiaolong and Hang, Mengyue and Liu, Xiaoyi and Zhou, Qinghai and Yang, Chaofei and Liu, Yiqun and Ruan, Yichen and Chen, Laming and Chen, Yuxin and others},
  booktitle={Proceedings of the 34th ACM International Conference on Information and Knowledge Management},
  pages={6225--6233},
  year={2025}
}

@inproceedings{zhou2018deep,
  title={Deep interest network for click-through rate prediction},
  author={Zhou, Guorui and Zhu, Xiaoqiang and Song, Chenru and Fan, Ying and Zhu, Han and Ma, Xiao and Yan, Yanghui and Jin, Junqi and Li, Han and Gai, Kun},
  booktitle={Proceedings of the 24th ACM SIGKDD international conference on knowledge discovery \& data mining},
  pages={1059--1068},
  year={2018}
}

@article{zhang2025onetrans,
  title={OneTrans: Unified Feature Interaction and Sequence Modeling with One Transformer in Industrial Recommender},
  author={Zhang, Zhaoqi and Pei, Haolei and Guo, Jun and Wang, Tianyu and Feng, Yufei and Sun, Hui and Liu, Shaowei and Sun, Aixin},
  journal={arXiv preprint arXiv:2510.26104},
  year={2025}
}

@inproceedings{han2025mtgr,
  title={Mtgr: Industrial-scale generative recommendation framework in meituan},
  author={Han, Ruidong and Yin, Bin and Chen, Shangyu and Jiang, He and Jiang, Fei and Li, Xiang and Ma, Chi and Huang, Mincong and Li, Xiaoguang and Jing, Chunzhen and others},
  booktitle={Proceedings of the 34th ACM International Conference on Information and Knowledge Management},
  pages={5731--5738},
  year={2025}
}

@article{deng2025onerec,
  title={Onerec: Unifying retrieve and rank with generative recommender and iterative preference alignment},
  author={Deng, Jiaxin and Wang, Shiyao and Cai, Kuo and Ren, Lejian and Hu, Qigen and Ding, Weifeng and Luo, Qiang and Zhou, Guorui},
  journal={arXiv preprint arXiv:2502.18965},
  year={2025}
}

@article{zhai2024actions,
  title={Actions speak louder than words: Trillion-parameter sequential transducers for generative recommendations},
  author={Zhai, Jiaqi and Liao, Lucy and Liu, Xing and Wang, Yueming and Li, Rui and Cao, Xuan and Gao, Leon and Gong, Zhaojie and Gu, Fangda and He, Michael and others},
  journal={arXiv preprint arXiv:2402.17152},
  year={2024}
}

@inproceedings{zhu2025rankmixer,
  title={Rankmixer: Scaling up ranking models in industrial recommenders},
  author={Zhu, Jie and Fan, Zhifang and Zhu, Xiaoxie and Jiang, Yuchen and Wang, Hangyu and Han, Xintian and Ding, Haoran and Wang, Xinmin and Zhao, Wenlin and Gong, Zhen and others},
  booktitle={Proceedings of the 34th ACM International Conference on Information and Knowledge Management},
  pages={6309--6316},
  year={2025}
}

@inproceedings{chai2025longer,
  title={Longer: Scaling up long sequence modeling in industrial recommenders},
  author={Chai, Zheng and Ren, Qin and Xiao, Xijun and Yang, Huizhi and Han, Bo and Zhang, Sijun and Chen, Di and Lu, Hui and Zhao, Wenlin and Yu, Lele and others},
  booktitle={Proceedings of the Nineteenth ACM Conference on Recommender Systems},
  pages={247--256},
  year={2025}
}

@inproceedings{xia2025transact,
  title={TransAct V2: Lifelong User Action Sequence Modeling on Pinterest Recommendation},
  author={Xia, Xue and Joshi, Saurabh and Rajesh, Kousik and Li, Kangnan and Lu, Yangyi and Pancha, Nikil and Badani, Dhruvil and Xu, Jiajing and Eksombatchai, Pong},
  booktitle={Proceedings of the 34th ACM International Conference on Information and Knowledge Management},
  pages={6881--6882},
  year={2025}
}

@article{ricci2021recommender,
  title={Recommender systems: Techniques, applications, and challenges},
  author={Ricci, Francesco and Rokach, Lior and Shapira, Bracha},
  journal={Recommender systems handbook},
  pages={1--35},
  year={2021},
  publisher={Springer}
}

@article{zhang2019deep,
  title={Deep learning based recommender system: A survey and new perspectives},
  author={Zhang, Shuai and Yao, Lina and Sun, Aixin and Tay, Yi},
  journal={ACM computing surveys (CSUR)},
  volume={52},
  number={1},
  pages={1--38},
  year={2019},
  publisher={ACM New York, NY, USA}
}

@inproceedings{li2024large,
  title={Large language models for generative recommendation: A survey and visionary discussions},
  author={Li, Lei and Zhang, Yongfeng and Liu, Dugang and Chen, Li},
  booktitle={Proceedings of the 2024 joint international conference on computational linguistics, language resources and evaluation (LREC-COLING 2024)},
  pages={10146--10159},
  year={2024}
}

@inproceedings{wang2025generative,
  title={Generative large recommendation models: Emerging trends in llms for recommendation},
  author={Wang, Hao and Guo, Wei and Zhang, Luankang and Chin, Jin Yao and Ye, Yufei and Guo, Huifeng and Liu, Yong and Lian, Defu and Tang, Ruiming and Chen, Enhong},
  booktitle={Companion Proceedings of the ACM on Web Conference 2025},
  pages={49--52},
  year={2025}
}

@article{lu2019vilbert,
  title={Vilbert: Pretraining task-agnostic visiolinguistic representations for vision-and-language tasks},
  author={Lu, Jiasen and Batra, Dhruv and Parikh, Devi and Lee, Stefan},
  journal={Advances in neural information processing systems},
  volume={32},
  year={2019}
}

@article{shazeer2019fast,
  title={Fast transformer decoding: One write-head is all you need},
  author={Shazeer, Noam},
  journal={arXiv preprint arXiv:1911.02150},
  year={2019}
}

@inproceedings{steck2018calibrated,
  title={Calibrated recommendations},
  author={Steck, Harald},
  booktitle={Proceedings of the 12th ACM conference on recommender systems},
  pages={154--162},
  year={2018}
}

@article{tay2020efficient,
  title   = {Efficient Transformers: A Survey},
  author  = {Tay, Yi and Dehghani, Mostafa and Bahri, Dara and Metzler, Donald},
  journal = {arXiv preprint arXiv:2009.06732},
  year    = {2020}
}

@inproceedings{he2014practical,
  title={Practical lessons from predicting clicks on ads at facebook},
  author={He, Xinran and Pan, Junfeng and Jin, Ou and Xu, Tianbing and Liu, Bo and Xu, Tao and Shi, Yanxin and Atallah, Antoine and Herbrich, Ralf and Bowers, Stuart and others},
  booktitle={Proceedings of the eighth international workshop on data mining for online advertising},
  pages={1--9},
  year={2014}
}

@inproceedings{geng2022recommendation,
  title={Recommendation as language processing (rlp): A unified pretrain, personalized prompt \& predict paradigm (p5)},
  author={Geng, Shijie and Liu, Shuchang and Fu, Zuohui and Ge, Yingqiang and Zhang, Yongfeng},
  booktitle={Proceedings of the 16th ACM conference on recommender systems},
  pages={299--315},
  year={2022}
}

@article{rajput2023recommender,
  title={Recommender systems with generative retrieval},
  author={Rajput, Shashank and Mehta, Nikhil and Singh, Anima and Hulikal Keshavan, Raghunandan and Vu, Trung and Heldt, Lukasz and Hong, Lichan and Tay, Yi and Tran, Vinh and Samost, Jonah and others},
  journal={Advances in Neural Information Processing Systems},
  volume={36},
  pages={10299--10315},
  year={2023}
}

@inproceedings{pi2020search,
  title={Search-based user interest modeling with lifelong sequential behavior data for click-through rate prediction},
  author={Pi, Qi and Zhou, Guorui and Zhang, Yujing and Wang, Zhe and Ren, Lejian and Fan, Ying and Zhu, Xiaoqiang and Gai, Kun},
  booktitle={Proceedings of the 29th ACM International Conference on Information \& Knowledge Management},
  pages={2685--2692},
  year={2020}
}

@article{ainslie2023gqa,
  title={Gqa: Training generalized multi-query transformer models from multi-head checkpoints},
  author={Ainslie, Joshua and Lee-Thorp, James and De Jong, Michiel and Zemlyanskiy, Yury and Lebr{\'o}n, Federico and Sanghai, Sumit},
  journal={arXiv preprint arXiv:2305.13245},
  year={2023}
}

@inproceedings{qin2020user,
  title={User behavior retrieval for click-through rate prediction},
  author={Qin, Jiarui and Zhang, Weinan and Wu, Xin and Jin, Jiarui and Fang, Yuchen and Yu, Yong},
  booktitle={Proceedings of the 43rd International ACM SIGIR Conference on Research and Development in Information Retrieval},
  pages={2347--2356},
  year={2020}
}

@article{dao2022flashattention,
  title={Flashattention: Fast and memory-efficient exact attention with io-awareness},
  author={Dao, Tri and Fu, Dan and Ermon, Stefano and Rudra, Atri and R{\'e}, Christopher},
  journal={Advances in neural information processing systems},
  volume={35},
  pages={16344--16359},
  year={2022}
}

\newpage 

\appendix
\label{sec:appendix}
\section{Complexity Analysis}
\label{sec:complexity-append}

Let $L_{\text{local}} \ll L$ be the size of the local attention window. For simplicity, assume the decoding head $f_{\text{dec}}(\cdot)$ is a linear function, and using causal attention for behavior stream encoding.

\paragraph{Training complexity.}
When the behavior encoder uses causal self-attention, the entire behavior stream can be encoded once in a single left-to-right pass with cost $\mathcal{O}\!\left(L^2 d + L d^2\right)$, which is amortized across all serving events for the user. After which each serving event incurs only event-level computation.
At each serving event, TransX applies grouped multi-query sparse cross attention
and parallel multi-action decoding, the per-event cost is $\mathcal{O}\!\left( m L_{\text{local}} d + m k d \right)$. 

Aggregated over $T$ serving events,
the total per-user training complexity of TransX is: $$\mathcal{O}\!\left(L^2 d + L d^2\right)+\mathcal{O}\!\left(T m L_{\text{local}}d+T m k d \right).$$
Crucially, the expensive sequence encoding term is independent of $T$,
and the remaining cost scales \emph{linearly} with the number of serving events.
In contrast, many sequential or generative recommendation models re-encode the
behavior prefix for each serving event. For a representative prefix re-encoding baseline with average prefix length $\bar L$, the per-user training cost is: $$\sum_{t=1}^{T} \mathcal{O}\!\left( \bar L^2 d + \bar L d^2 \right) = \mathcal{O}\!\left( T (\bar L^2 d + \bar L d^2) + Tm k d \right).$$

\paragraph{Online computation complexity.}
During online inference, TransX avoids any Transformer-layer computation over the
full behavior sequence.
Behavior representations and their corresponding key--value tensors are computed
offline or nearline and retrieved from cache.
As a result, online computation consists only of event-level operations.
For each serving event, candidate-to-query encoding incurs
$\mathcal{O}(m d^2)$ operations.
Grouped multi-query sparse cross attention reuses shared behavior-derived keys and
values across all candidates and restricts attention to $L_{\text{local}}$ recent
behavior tokens and a constant number of global summary tokens, yielding
$\mathcal{O}\!\left(m L_{\text{local}} d\right)$ operations.
Parallel non-autoregressive action decoding produces action logits for all candidates
using shared action token embeddings, with cost $\mathcal{O}(m k d)$.
The total online computation per serving event is therefore: $$\mathcal{O}\!\left(m d^2+m L_{\text{local}} d+m k d\right),$$
which scales linearly with the number of candidates and is independent of the total behavior sequence length $L$. In contrast, many sequential recommendation and GR models apply full or dense cross attention between candidates and the entire behavior sequence, leading to online complexity $\mathcal{O}(m L d^2)$ or higher. 

\paragraph{Online memory complexity.} The online memory footprint of TransX is similarly bounded.
For each request, both the on-disk behavior sequence encoding cache and in-memory KV cache are from only $L_{\text{local}}$ recent behavior tokens and a global  token, with memory cost of $\mathcal{O}(L_{\text{local}} d)$.

\section{Action Prediction Head in Decoder}
\label{sec:actionhead-append}

The modular design of TransX enables the reuse of any mature, production-validated prediction heads while cleanly separating sequence modeling from downstream action prediction.
As a simple instantiation, the decoder-side action prediction head can adopt a Deep \& Wide architecture. In this design, the deep component incorporates the candidate-conditioned cross-attention representation together with learned user and item embeddings, while the linear component captures numerical features and low-order feature interactions. Formally, the decoder head is defined as
\begin{equation*}
\begin{split}
    & f_{\text{dec}}(\mathbf{z}_t^{(u,c)}, x_t^{(u,c)}) \\
    & =\text{concat}([\text{linear part of } x_t^{(u,c)}, \text{FFN}(\mathbf{z}_t^{(u,c)}, \text{deep part of } x_t^{(u,c)})]),
\end{split}
\end{equation*}
where $\mathbf{z}^{(u,c)}$ is the cross attention output for candidate $c$ with user $u$'s behavior sequence up to time $t$.
All candidates can be scored fully in parallel during online serving.

The more expressive action prediction head used in industry today often adopts a Transformer-based architecture. In this setting, candidate features, user features, and contextual representations are projected into key--value pairs and combined through dot-product self-attention to model higher-order feature interactions. For a serving event $\mathcal{E}_t$ with $m$ candidates, the input token sequence to the action prediction head is constructed as:
$$S^{(\mathcal{E})}_t:=[x_t^{\text{context}};x_t^{\text{user}};[x_t^{c_1},\mathbf{z}_t^{(u,c_1)}];\ldots;[x_t^{c_m},\mathbf{z}_t^{(u,c_m)}]],$$
where $x_t^{\text{context}}$ and $x_t^{\text{user}}$ denote event-level context and user features, and $x_t^{c}$ denotes candidate-specific features. A customized attention mask is applied such that each candidate token attends only to the context and user tokens, while remaining independent of other candidates (as illustrated in the top right corner of Figure~\ref{fig:system}.). Under this masking scheme, applying a Transformer block to $\mathbf{S}_t^{(\mathcal{E})}$ computes the decoder function in parallel for each candidate:
\begin{equation*}
\begin{split}
    & f_{\text{dec}}(\mathbf{z}_t^{(u,c)}, x_t^{(u,c)}) \\
    & =\text{Transformer}([x_t^{\text{context}};x_t^{\text{user}};[x_t^{\text{candidate}},\mathbf{z}_t^{(u,c)}]),
\end{split}
\end{equation*}
where $x_t^{\text{context}}$, $x_t^{\text{user}}$, $x_t^{\text{candidate}}$ represents the context features, user features, and candidate-specific features (including pair features) in $x_t^{(u,c)}$. 

Finally, techniques proposed in HiFormer \cite{gui2023hiformer} and OneTrans \cite{zhang2025onetrans} can be naturally incorporated into this attention-based prediction head to further improve the handling of heterogeneous feature types.

\section{Additional Model Benchmarking Details}
\label{sec:benchmark-append}

We provide additional details for our offline benchmarking experiments:
\begin{itemize}[leftmargin=*]
    \item All models in our offline benchmarking experiments are trained using the Adam optimizer with an initial learning rate of 0.001. We adopt a learning-rate schedule consisting of linear warmup followed by cosine decay. Model selection is performed based on validation AUC with early stopping, and all reported offline metrics are computed on the held-out test set.
    \item All models are implemented with TensorFlow 2.11 library and trained on an on-premise NVIDIA H100 GPU cluster under identical hardware and software configurations to ensure fair comparison.
    \item Across all models, including discriminative baselines, action-transduction models, and next-action generation models, we use the same binary cross-entropy loss for click and no-click prediction. For models that predict action distributions, the binary loss is applied consistently to the click action to align evaluation objectives across architectures.
    \item To control model capacity, we fix the number of Transformer blocks to two for all Transformer-based baselines and TransX. Model width, including projection and FFN hidden dimensions, is adjusted to approximately match MFLOPs across models. After capacity matching, dropout rate is treated as the primary hyperparameter and tuned on the validation set.
    \item Unless otherwise stated, all models are trained with the same batch size, sequence construction and tokenization, user / item / contextual features, and data preprocessing pipelines, ensuring that performance differences are attributable to architectural design rather than training or data artifacts.
\end{itemize}

\begin{table}[htb]
\centering
    \begin{tabular}{c|ccc|ccc}
    \hline
    Attention mask & \multicolumn{3}{c|}{Causal} & \multicolumn{3}{c}{Bidirectional} \\ 
    \# Layers & h=1 & h=2 & h=3 & h=1 & h=2 & h=3 \\ \hline
    No pos bias & 0.850 & .856 & 0.859 & 0.852 & 0.859 & 0.861  \\
    Abs pos bias & 0.851 & .858 & 0.860 & 0.853 & 0.861 & 0.863  \\
    \textbf{Rel pos bias} & \textbf{0.852} & \textbf{0.862} & \textbf{0.863} & \textbf{0.853} & \textbf{0.863} & \textbf{0.864} \\ \hline
    \end{tabular}
\caption{\small Ablation study for the encoder. Reported results are AUC. Here, $h$ denotes the the number of Transformer layers, \texttt{abs pos bias} refers to learnt absolute position encoding \cite{devlin2019bert}, \texttt{rel pos bias} refers to relative position bias \cite{dai2019transformer}.}
\label{tab:encoder-ablation}
\end{table}
\vspace{-0.3cm}

\begin{table}[hbt]
    \centering
    \begin{tabular}{c|cccc}
    \hline
         & $L_{\text{local}}=5$ & $L_{\text{local}}=10$ & $L_{\text{local}}=15$ & $L_{\text{local}}=20$   \\ \hline
        No global token & 0.853  & 0.856  & 0.858  & 0.859  \\ 
        Mean/Max pool & 0.856  & 0.860  & 0.861  & 0.862   \\ 
        \textbf{Attention pool} & \textbf{0.857} & \textbf{0.862}  & \textbf{0.864}  &  \textbf{0.865}  \\ \hline
    \end{tabular}
    \caption{\small Ablation study for the sparse cross attention. Reported results are AUC. \emph{No global token} disables global aggregation, \emph{Mean/Max pool} uses the better of mean or max pooling to form a global token, and \emph{Attention pool} learns a weighted global representation.}
    \label{tab:cross-attn-ablation}
\end{table}
\vspace{-0.3cm}

\section{Further deployment details}
\label{sec:coldstart-append}

\paragraph{Handling cold-start recommendation} 
We adopt a pragmatic system-level solution to handle users with no observable behavior history. The approach consists of the following components:
\begin{itemize}[leftmargin=*]
    \item \textbf{Offline training.} We train a lightweight backup model on cold-start traffic that excludes the behavior-stream encoder and cross-attention components, relying solely on event-level, user, and candidate features.
    \item \textbf{Online serving.} During inference, the serving backend first attempts to retrieve the cached behavior-stream encoding for the requesting user. If no cache entry is found (e.g., for new or returning users), the system automatically falls back to the backup model and serves recommendations using only event-level features. 
\end{itemize}
In our production use case, approximately $2\%$ of the total traffic were routed to the backup model.

For users with short behavior histories (e.g., fewer than 10 interactions), we do not apply a separate handling strategy in order to reduce system complexity. Since TransX retrieves behavior representations only within a localized context window, its performance is less sensitive to sequence-length skew, allowing the full model to be applied without special-case logic.

In online A/B testing, we observe neutral CTR changes for users with no eligible behavior history, and statistically significant CTR improvements for users with behavior sequences shorter than $L_{\text{local}}$.

\paragraph{Cache management} 
Our system adopts the following strategies to achieve effective cache management:
\begin{itemize}[leftmargin=*]
    \item \textbf{Warm start.} During initial model launch, for members with eligible behaivor sequence, we pre-compute all user's behaivor encoding representations and persist them in on-disk cache. For newly active users, behavior representations are incrementally materialized and cached upon the first few serving events with positive engagement.
    \item \textbf{Lifecycle.} Cached representations are stored with a time-to-live (TTL) and evicted based on recency of access, ensuring that memory usage scales with the set of active users rather than total user population. 
    \item \textbf{Versioning and consistency.} Cached entries are versioned by both the model identifier and the feature schema version. Upon model upgrades or feature changes, stale cache entries are automatically invalidated, preventing mismatches between cached representations and the active serving model.
    \item \textbf{Failure recovery.} In the event of cache corruption, eviction spikes, or transient storage failures, the serving system will fall back to the cold-start backup model and resume on-the-fly behavior encoding.
\end{itemize}

\paragraph{Nearline update}
To keep behavior-stream representations fresh while meeting latency and throughput constraints, the system adopts two complementary nearline update mechanisms using \emph{event-based} and \emph{time-based} triggers, depending on data availability and system requirements.
\begin{itemize}[leftmargin=*]
    \item for highly active users, we adopt a Kafka-based streaming pipeline that continuously ingests user interaction events. Behavior updates are processed / encoded incrementally, and used to refresh the cached behavior-stream encodings in near real time. This setup enables low-latency incorporation of new user actions while amortizing computation across events.
    \item for other users, we employ an HDFS-based batch pipeline. In this mode, user behavior events are periodically aggregated over fixed time windows and used to recompute behavior-stream encodings in bulk.
\end{itemize}
Both mechanisms update the same versioned cache entries to guarantee completeness and consistency over longer horizons. 

\begin{remarkbox}
\textbf{Remark: incremental encoding}. In the streaming-based incremental update setting, \emph{we do not re-encode the entire behavior sequence upon the arrival of a new interaction}. Instead, when using a causal Transformer encoder with relative positional bias \cite{dai2019transformer}, the cached representations of historical tokens remain unchanged, and only the representation of the newly appended behavior token is computed. This incremental update mirrors standard key--value caching in autoregressive Transformers and allows nearline updates to scale with the number of new events rather than total sequence length.
\end{remarkbox}

\section{Additional Online A/B Experiment Configuration and Results}
\label{sec:additional-AB}

In this section, we provide additional details on A/B experiment configuration, stability, and segmented analysis to further validate the robustness and scalability of TransX in production.



\paragraph{A/B results segmented by user activity level}
We segment users based on quantiles of their 1-week pre-experiment activity counts into \texttt{low}, \texttt{medium}, and \texttt{high} activity groups. Users with no prior activity are categorized as \texttt{cold-start}.
As shown in Table \ref{tab:activity-segmented}, TransX achieves consistent gains across all active user segments, with larger improvements for higher-activity users (which is a standard observation for industrial sequential recommendation models). For cold-start users, performance remains neutral (not statistically significant), indicating that TransX does not degrade performance when behavioral signals are absent.

\begin{table}[htb]
    \centering
    \begin{tabular}{c|c|c|c|c}
    \hline 
       Activity level  & High act. & Medium act. & Low act. & Cold-start \\
       Traffic\%  & 32\% & 48\% & 16\% & 4\% \\ \hline \hline
       CTR lift & \textcolor{black}{+6.6\% $\uparrow$}  & \textcolor{black}{+5.5\% $\uparrow$}  & \textcolor{black}{+1.8\% $\uparrow$} & +0.01 \\ \hline 
       CVR lift & \textcolor{black}{+4.7\% $\uparrow$} & \textcolor{black}{+4.3\% $\uparrow$} & \textcolor{black}{+3.6\% $\uparrow$} & -0.01 \\ \hline \hline
    \end{tabular}
    \caption{Online performance segmented by user activity level (according to treatment population's 1-week pre-experiment activity counts).}
    \label{tab:activity-segmented}
\end{table}

\paragraph{A/B results segmented by history length}
We further segment users based on quantiles of their 180-day behavior sequence length into \texttt{long}, \texttt{medium}, and \texttt{short} history groups. Users without valid behavior sequences are categorized as \texttt{no history}.
Table \ref{tab:history-segmented} shows that TransX delivers statistically significant improvements across all non-empty history buckets. The treatment effects are higher for users with longer history lengths -- this trend reflects TransX’s ability to better leverage long-term behavioral signals. Performance remains neutral for users without history, consistent with the activity-based analysis.

\begin{table}[hbt]
    \centering
    \begin{tabular}{c|c|c|c|c}
    \hline \hline
       History len.  & Long & Medium & Short & No history \\ \hline
       p50 len.  & 163 & 67 & 13 & 0 \\ \hline \hline
       CTR lift & \textcolor{black}{+6.2\% $\uparrow$} & \textcolor{black}{+6.0\% $\uparrow$} & \textcolor{black}{+5.5\% $\uparrow$} & 0.0 \\ \hline 
       CVR lift & \textcolor{black}{+4.8\% $\uparrow$} & \textcolor{black}{+4.3\% $\uparrow$} & \textcolor{black}{+4.1\% $\uparrow$} & 0.0 \\ \hline \hline
    \end{tabular}
    \caption{Online performance segmented by user behavior history length (in past 180 days). The p50 length row provide the medium sequence length of the segments.}
    \label{tab:history-segmented}
\end{table}

\paragraph{Long-tail serving analysis}
We analyze serving latency under different candidate set sizes to evaluate scalability in heavy-tailed traffic scenarios. Table \ref{tab:latency-breakdown} shows:
\begin{itemize}
    \item amortized encoding significantly reduces latency and achieving near-linear scaling,
    \item KV caching further stabilizes latency growth, 
    \item the reduction in latency is more pronounced for larger candidate set size.
\end{itemize}

\begin{table}[htb]
    \centering
    \begin{tabular}{c|c|c|c|c}
    \hline \hline
    Candidate size  & 100-300 & 300-500 & 500-1,000 & 1,000+ \\
    Traffic\%  & 22\% & 26\& & 35\% & 17\% \\ \hline \hline
    raw p99 latency & 201ms  & 255ms  & 289ms  &  330ms   \\
    w. amortized encoding  &  60ms &  71ms & 86ms  & 113ms   \\
     and caching &   &   &   &    \\
    + KV caching &  43ms & 49ms  & 58ms  &  74ms  \\ \hline \hline
    \end{tabular}
    \caption{p99 model inference latency (excluding upstream retrieval, networking, and post-scoring processing) under different candidate set sizes. The Traffic\% row provides the proportion of the online traffic that falls into the candidate size bucket.}
    \label{tab:latency-breakdown}
\end{table}



\paragraph{Overall Takeaway.} TransX's online improvements are stable over time (5-week evaluation) and consistent across user segments, with larger gains for high-activity / long-history users and no regression in cold-start scenarios. Importantly, these gains hold under realistic serving conditions, including heavy-tailed candidate sizes and delayed cache refresh, demonstrating that TransX is not sensitive to infrastructure imperfections. At the same time, the system achieves substantial latency reductions (up to ~80\% p99) while lifting performance, confirming that the model–infrastructure co-design effectively advances the accuracy–efficiency Pareto frontier. Together, these results show that TransX’s benefits are not confined to specific slices or ideal conditions, but generalize reliably across user populations and production constraints.

\end{document}